\documentclass[twocolumn, switch]{article} 

\usepackage{preprint}

\usepackage{amsmath, amsthm, amssymb, amsfonts}

\usepackage[numbers,square]{natbib}

\usepackage[utf8]{inputenc}	
\usepackage[T1]{fontenc}	
\usepackage{xcolor}		
\usepackage[colorlinks = true,
            linkcolor = purple,
            urlcolor  = blue,
            citecolor = cyan,
            anchorcolor = black]{hyperref}	
\usepackage{booktabs} 		
\usepackage{nicefrac}		
\usepackage{microtype}		
\usepackage{lineno}		
\usepackage{float}			
\usepackage{multicol}		
\usepackage{multirow}

\usepackage{newfloat}
\DeclareFloatingEnvironment[name={Supplementary Figure}]{suppfigure}
\usepackage{sidecap}
\sidecaptionvpos{figure}{c}

\usepackage{titlesec}
\titlespacing\section{0pt}{12pt plus 3pt minus 3pt}{1pt plus 1pt minus 1pt}
\titlespacing\subsection{0pt}{10pt plus 3pt minus 3pt}{1pt plus 1pt minus 1pt}
\titlespacing\subsubsection{0pt}{8pt plus 3pt minus 3pt}{1pt plus 1pt minus 1pt}

\usepackage{tikz,xcolor,hyperref}

\definecolor{lime}{HTML}{A6CE39}
\DeclareRobustCommand{\orcidicon}{
	\begin{tikzpicture}
	\draw[lime, fill=lime] (0,0) 
	circle [radius=0.16] 
	node[white] {{\fontfamily{qag}\selectfont \tiny ID}};
	\draw[white, fill=white] (-0.0625,0.095) 
	circle [radius=0.007];
	\end{tikzpicture}
	\hspace{-2mm}
}
\foreach \x in {A, ..., Z}{\expandafter\xdef\csname orcid\x\endcsname{\noexpand\href{https://orcid.org/\csname orcidauthor\x\endcsname}
			{\noexpand\orcidicon}}
}

\title{Binary-level Software Compatibility Tool Agreement}

\usepackage{xwatermark}
\newwatermark[firstpage,color=gray!60,angle=90,scale=0.32, xpos=-4.05in,ypos=0]{\href{https://doi.org/}{\color{gray}{}}}
\newwatermark[firstpage,color=gray!60,angle=90,scale=0.32, xpos=3.9in,ypos=0]{\href{https://doi.org/}{\color{gray}{Preprint doi}}}
\newwatermark[firstpage,color=gray!90,angle=0,scale=0.28, xpos=0in,ypos=-5in]{*correspondence: \texttt{sochat1@llnl.gov}}

\usepackage{authblk}

\author[1]{Vanessa Sochat\orcidA{}}
\author[2]{Tim Haines\orcidB{}}

\affil[1]{Lawrence Livermore National Lab, Livermore CA, USA}
\affil[2]{University of Wisconsin Madison, Madison WI, USA}

\usepackage{listings}
\usepackage{color}

\definecolor{dkgreen}{rgb}{0,0.6,0}
\definecolor{gray}{rgb}{0.5,0.5,0.5}
\definecolor{mauve}{rgb}{0.58,0,0.82}

\begin{document}

\twocolumn[ 
  \begin{@twocolumnfalse} 
  
\maketitle

\begin{abstract}
Application Binary Interface (ABI) compatibility is essential for system or software updates to ensure that libraries continue to function. Tools that can assess a binary or library ABI can thus be used to make predictions about compatibility, and predict downstream bugs by informing developers and users about issues. In this work, we are interested in describing a set of well-known tools for assessing ABI, and testing them in a controlled set experiments to assess tool agreement. We run 7660 smaller experiments across tools (N=30,640 total results) to evaluate not only predictions, but also each tool's ability to provide detail about underlying issues. In this work, along with highlighting the problem of assessing ABI compatibility and critiquing the pros and cons of currently available tools, we provide guidance to developers interested to test ABI based on our empirical results and suggestions for future work.
\end{abstract}


\vspace{0.35cm}

  \end{@twocolumnfalse} 
] 

\section{Introduction}
\label{sec:introduction}
\label{sec:abi}

In the same way that an Application Programming Interface (API) provides an interface to access the functionality and data provided by a library and a calling client must understand the details of these interfaces to make successful requests, the Application Binary Interface (ABI) of any software or operating system is interested in similar compatibility, but on the level of machine code. This means that as new software versions are released to update a system, the functions, parameter types, and many other attributes of the new binaries must be compatible with other libraries on the system. Akin to calling a RESTful API endpoint that expects a particular set of parameters from a client, if you change or otherwise provide an incorrect set, your client will no longer work with the API. The same is true for the binaries on a system, and careful work must be taken to ensure enduring compatibility.  Any new library installed to the system must function without requiring recompilation of other libraries already in the ecosystem. 

Ensuring compatibility is a non-trivial task, as different systems provide different compilers, operating systems, and supporting libraries, and these varying tool chains are expected to build and run the software. Even a simple case of a function changing argument types can mean a break in functionality. Thus, introducing multiple compiler tool-chains, library versions, languages, and even architectures for connected systems \cite{hpc-arches}, presents the community with an enduring challenge: How do we assess ABI compatibility, and intelligently update tools with some assurance that the software will continue to work? 

\subsection{The Origins of ABI Compatibility}

When we install software on a computer and run it, we often take it for granted that it just works. Software is created by way of the compilation process - starting with source code, and compiling it with a language-specific compiler into byte-code that a particular machine can execute. The files containing these byte code can either be referred to as executables or binary libraries (often just called libraries). Installation of these binaries then comes down to either retrieving a previously built artifact (e.g., from a package manager) or building from source. The ultimate successful outcome of this process is the software running successfully on your system.

The question about ABI compatibility dates back to the origins of software. Although we cannot pin an exact date on when people started to think about compatibility, we can go as far back to 1951 and find that engineers were thinking about a documenting “Specifications of Library Subroutines” \cite{wilkes1951preparation} that detailed the same concerns that we have today -- namely types, addresses, and functionality. It was not until the 1980’s that discussion started about application binary standards, and how ABI was important for microprocessors and architecture compatibility \cite{1988infoworld,IDG_Network_World_Inc1987-vu} across computer magazines \cite{IDG_Enterprise1989-sk,InfoWorld_Media_Group1989-ok}, academic publications \cite{national1989ncga,1988computer,1989edp}, and even user group newsletters alike \cite{Auug1988-be}. More specifically, the Unix System V version I was first released in 1983 \footnote{\url{https://en.wikipedia.org/wiki/UNIX\_System\_V}}, and was a prominent standard under discussion for that decade \cite{IDG_Enterprise1989-sk,1988mini}. At the time of this writing (October, 2022) there are thousands of mentions \cite{search}, and we can speculate that this growth reflects the term becoming more understood and used across the software development community. For the interested reader, we include a brief list of notable ABI initiatives in Table \ref{tab:table1}.

\begin{table*}
	\caption{\label{tab:table1}Early Application Binary Interface Efforts}
	\centering
	\begin{tabular}{lll}
		\toprule
		\cmidrule(r){1-3}
		Name     & Published     & Year \\
		\midrule
		UUABI "Unix to Unix`` ABI \cite{ieee1989digest} & Compcon  & 1989 \\
		PAX "Parallel Architecture Extended for parallel computing`` \cite{predicasts1989predicasts} & Penn State & 1989  \\
		Unix System V \footnotemark[2] &       & 1983\\
		Intel Binary Compatibility  \footnote{https://en.wikipedia.org/wiki/Intel\_Binary\_Compatibility\_Standard} & Intel / AT\&T, SCO & 1988\\
    \bottomrule
	\end{tabular}
\end{table*}

This need for compatibility, and discussing it, has led to the creation of communities of people that are concerned with software’s ABI, and standards documents \cite{noauthor_undated-ke} that define strict rules for how software in a particular language is expected to interact with a specific architecture. More abstractly, these standards define a contract between programs that promises consistent interoperability between applications even as components are updated \cite{noauthor_undated-da}. ABI compatibility also makes a promise that, given that two binaries follow an ABI specification, we can move one binary to another system that conforms to the same system without any modifications \cite{7168530}. Any given library on a system is considered ABI compatible if the interface that it provides to access its functions and data is compatible with other libraries on that system that need it. However, in practice, the idea of ABI compatibility goes well beyond what is written down in a standards document, and includes any detail about the structure or functionality of a system that might render it incompatible with another system’s interface. 

\subsection{Problems that warrant understanding ABI}
\label{sec:problems}

While there is no universally agreed upon definition for what features constitute an ABI, there is much discussion around what people think an ABI is \cite{7168530,5952657} and these ideas are generally based around real problems that developers face when building software. Questions and topics include (but are not limited to) parameter passing in registers, data types, exceptions, global variables, configuration, optimization, dynamic dispatching, and virtual tables in C++. Common compatibility problems faced by developers include library versions and change over time, and implementation and compatibility problems in compilers. We direct the interested reader to the provided references for a more comprehensive listing of ABI problems. An aggregation of these ideas might be considered as a larger interface for ABI. Many of these problems (e.g., configuration- and optimization-related problems), while important to be worked on, have a level of complexity that deems them out of scope for this work.

The simplest, and most consistent challenge for a library ABI is with respect to the interfaces that libraries provide. As an example, if a math library version 1.0.0 is compiled with a function to do addition that expects two integer arguments, if the library changes those integer arguments to floats for version 2.0.0, a previous library that worked with 1.0.0 and expects integers will not work with version 2.0.0. Running such a library relies on dynamic linking \cite{10.1145/2786763.2694392}, a process by which the function signatures "symbols" are loaded into memory and shared. While outside the scope of this paper, we have provided a comprehensive description of the dynamic linking process for the interested reader \cite{noauthor_undated-om}. In the case of our C++ example here, this means that changing the function name or parameter names or types changes a symbol name. It follows that removing a function deletes a symbol, and adding a function adds a new symbol. An example of function signatures and the C++ symbols called "mangled strings" they produce are shown below, before and after changing a parameter type:

\label{example:function-parameter-add}
\begin{lstlisting}
int Add(int a, int b);          // _ZN11MathLibrary10Arithmetic3AddEii 
int Add(double a, double b);   // _ZN11MathLibrary10Arithmetic3AddEdd 
\end{lstlisting}

Notice that the mangled strings end with two different sets of letters that indicate parameter types, "ii" to indicate two integer parameters, and "dd" to indicate two double parameters, respectively. A program that uses this math library that is expecting the first symbol (integer parameters to "Add") would fail if provided the second library where they have been changed to doubles due to the missing symbol. This reflects a simple example of a set of much more complex issues around dynamic linking. While it allows for patching individual libraries to address security issues or add functionality without needing to rebuild an entire system, any change in a library could have dire consequences if the new library is not compatible with other software on the system. This particular problem is exacerbated for other languages that do not use parameter type information in the mangled strings. As an example, the same library compiled with C would produce the symbol "Add" in both cases, meaning that the incompatibility could not be detected at link time. A more complete example of this math library \cite{noauthor_undated-gt} with releases is provided.

Having these problems has led to the development of tools that try to solve them. At the highest level,  much of ABI compatibility is the responsibility of a compiler implementation, and thus falls under the responsibility of compiler authors. While compilers can catch many of these issues at compile time, third party tooling can assist with detection after this step. Any such detection tool inherently has an underlying model about what constitutes compatibility. For the purposes of this work, we can define a model of ABI compatibility as an approach a prediction tool makes about ABI compatibility. When two binaries are determined to not be compatible, we call this a \textit{breakage} or \textit{ABI break}. The model may operate in different contexts or environments, such as doing a pairwise comparison of two libraries, versus a global assessment of compatibility of an embedded system with a host architecture. Having such tools, or more generally checking for ABI compatibility, is important when performing code instrumentation to ensure memory safety \cite{Burow_2018} or when improving compilers for ABI-sensitive regions of code generation \cite{https://doi.org/10.48550/arxiv.2112.01397}.

Thus, a reasonable lens to look through to understand our current ability to assess ABI is through the tooling. While we cannot assess the accuracy of tool predictions without a ground truth for whether two binaries are compatible, we can compare consistency across tools and uncover interesting cases of disagreement. We can highlight these areas as important for future, more in-depth work, and to provide suggestions to developers for good practices for developing their software. For this work, we will outline existing models and tools for assessing ABI \ref{sec:methods}, design an experiment that tests tools across several versions of an operating system \ref{sec:experiment}, discuss results, pros and cons of each tool, and provide suggestions for utilization \ref{sec:discussion}, and conclude by discussing future work in this space.

\section{Methods}
\label{sec:methods}

\subsection{Tools}
\label{sec:tools}

The tools discussed are scoped to those that work with the Executable and Linking Format (ELF) \cite{elf} and debugging information (DWARF) that is supported for it \cite{noauthor_undated-if}. These binary formats are the most widely used (Libabigail) and popular \cite{Hemel_undated-pr} across Unix compatible operating systems. To the knowledge of these authors, there are no well-established ABI checking tools for other platforms. These tools generally use language level models, meaning they look at variable types, sizes,
and other associated metadata. In these sections we will describe three tools that can be use to make an assessment, including libraries Libabigail and the ABI Compliance Checker, and a general method to assess symbols. In this paper we refer to these tools and stratgegy as \textit{libagigail}, \textit{abi-compliance-checker}, and \textit{symbols} in italics, respectively.

\subsection{Libabigail}
\label{sec:libabigail}

Libabigail is a project maintained by RedHat that was first publicly released in 2015 \cite{noauthor_undated-dk}. The C++ library tackles the problem of assessing ABI compatibility by reducing a library to a set of ABI artifacts called a “corpus” that can include types, functions and declarations, variables, and other metadata parsed from ELF and DWARF \cite{noauthor_undated-cp}. The library provides command line executables to generate these artifacts in XML format to work with tools to assess compatibility (abicompat), differences (abidiff), and output of the XML structures (abidw). The strongest use case for Libabigail is for Red Hat Enterprise Linux (RHEL), Fedora, and the Android operating system. Using Libabigail, developers ensure that they don't unintentionally break ABI when doing updates. While Libabigail uses DWARF to improve comparison, a strength of the tool is that it does not require it for doing a basic comparison using only symbols \cite{abidiff}.

\subsection{ABI Compliance Checker}
\label{sec:abilab}

The ABI Compliance Checker \cite{5952657,7168530,noauthor_undated-jk}, akin to Libabigail, is concerned with assessing ABI compatibility of C and C++ libraries. It does this by way of defining a set of rules that define ABI compatibility \cite{Ponomarenko_undated-ge} and also outputting XML to then compare against the rules. The authors make results of their checks available in a public web interface \cite{noauthor_undated-oh}. The comparisons are driven by a tool to dump out a representation of the binary (akin to a corpus) using the abi-dumper \cite{Ponomarenko_undated-wv} and then comparing two dumps with the abi-compliance-checker \cite{Ponomarenko_undated-yn}. Unlike Libabigail, the ABI compliance checker requires DWARF information to be present. These tools are appealing to developers as they come with visualization tools \cite{Ponomarenko_undated-si} to generate entire web interfaces for results.

\subsection{Symbols}
\label{sec:symbols}

While there are many tools for extracting symbols (e.g., dpkg-gensymbols, chkshlib, cmpdylib, cmpshlib), our choice of Python for running experiments led us to use the pyelfutils \cite{Bendersky_undated-aa} library to extract lists of symbols from binaries. For our compatibility assessment, we start with the ELF file's .symtab and .dynsym sections, filter down symbols to with size greater than 0, not belonging to an undefined section, not missing a type, and not local \cite{symbol-table}. An \textit{exported} symbol is one that is provided by a library and be consumed by some other library or executable binary via linking. If we let $A$ be an initial version of a library and $B$ a second, more recent, version of the same library, then \textit{symbols} calculates the simple set difference:

\begin{lstlisting}
missing-previously-found-exports(A, B):
    {exported A} \ {exported B}
\end{lstlisting}

In the above, an empty set would indicate that there is no change (and thus no ABI break) between the sets. This simple symbol comparison "symbols`` serves as a meaningful base case test. For the interested reader, the above functionality is implemented in the library "elfcall`` as the \textit{symbols} predictor \cite{noauthor_undated-om}.

\section{Experiment}
\label{sec:experiment}

Our experiment is concerned with tool agreement. Given the three predictors (Libabigail, ABI Laboratory, \textit{symbols}) described in Section \ref{sec:tools} run across a large set of pairwise comparisons of matched libraries (discussed below) between Fedora releases 34, 35, 36, and 37 we asked the question what is the agreement between the tools that an ABI break is detected. We used \textbf{abidiff} for Libabigail, the \textbf{abi-dumper} and \textbf{abi-compliance-tester} for the ABI Compliance checker, and the set comparison previously described for \textit{symbols}. With results from the predictors, we are interested in the agreement between predictors. Given the console output from each predictor, we will be able to classify predicted breakages into the categories of removals, changes, and additions for each of function parameters, virtual tables, enumerators and global variables.

\subsection{Reproducibility Statement}
\label{sec:reproducibility}

We approach tackling this problem with a goal of reproducibility - it would not be enough to run experiments that rely on our proprietary systems or software, so we do the entire analysis using publicly available continuous integration (CI), unit tests, and automated builds of containers with open source software. In addition, given the complexity of needing to provide several builds of prediction software that fold into common software to perform the experiments, the authors provide automated builds of all container bases required for the experiments. The containers are provided as automated builds alongside the main repository that outlines the experiment and how to run it \cite{noauthor_undated-px}, and the automated runs were done on a separate repository \cite{noauthor_undated-eg} that saves results as artifacts that are retrieved with the final analysis repository \cite{noauthor_undated-km}. All of our tooling, unit tests, and testing infrastructure are publicly available for others to reproduce or extend our work.

\subsection{Design}
\label{sec:design}

We start with pre-built containers across four Fedora versions with all predictors installed. Since we need to compare across versions, this requires saving an artifact from each Fedora version that includes all binaries and debug information directories, which are installed separately \cite{noauthor_undated-mg}. We can then create a second job in our continuous integration environment \ref{sec:reproducibility} that creates a matrix of pairwise Fedora versions (not comparing a version to itself) to run the analysis described previously. In layman's terms, we are asking each tool if there are ABI breaks between "equivalent`` libraries provided in different Fedora operating systems. We emphasize "equivalent'' because our best effort to match libraries means checking for equivalent directory and file name prefixes (e.g., /lib/dirx/libx.so.1 would be matched to /lib/dirx/libx.so.2 by way of the prefix "libx" and a matching absolute path parent directory name "/lib/dirx"). This means that it is possible to have name collisions, or cases of finding the same prefix for two different underlying libraries (e.g., as is common with symbolic links). However, we did not find any of these cases in our analysis. For each match, the comparison procedure is performed with the respective tools, essentially asking if the two versions are compatible. 

To mimic an upgrade process, for the directionality of our comparison we choose to compare the lower Fedora version (e.g., 34) with a newer one (e.g., 35). This asks the question if the newer library is ABI compatible with the older one. In addition to recording complete terminal output to later derive rationale for a tool's decision, we also record the library size (in bytes) along with the time to reach a decision to make an assessment about bytes/second and relative tool speed.

We developed a Python library, "spliced" \cite{noauthor_undated-wb} that runs the experiments within container bases matched to our Fedora versions, and with all predictors installed. Spliced serves as a wrapper to provide a set of binaries, A and B, to each of the predictors, to capture the output, and save to a common format (JSON) for further analysis. For each result (a comparison between A and B), we capture high level metadata about the binaries, sizes in bytes, and a set of predictions that are namespaced by the predictor. As these experiments are run in GitHub actions, the results themselves are saved as GitHub artifacts that can be programmatically retrieved for further analysis \cite{noauthor_undated-nl}.
  
\section{Results}
\label{sec:results}

The goal of our experiment is to do qualitative comparisons between the ABI compatibility assessment tools Libabigail, the ABI Laboratory, and \textit{symbols} (see Section \ref{sec:tools}). We ran pairwise comparisons of matching binaries between four versions of the Fedora operating system (34, 35, 36, 37), and use the output (JSON) from these runs to drive our exploratory analysis.

During our experiments, we often found that the abi-compliance-checker predictor was terminated prematurely by the Linux stack-smashing detector \cite{hulk-smash}. Attempting to re-run those analyses yielded inconsistent results such that we could not feasibly execute this predictor on all of the libraries under consideration. To account for this, we split the most recent set of results into two categories: those for which there are results for all three predictors and those for which there are results only for Libabigail and \textit{symbols}. The subset having all three predictors has $59.5^{71.0}_{43.8}\%$ fewer binaries \footnote{Throughout, we use the notation $average^{max}_{min}$ to represent the sample distribution of a quantity measured across all OS comparisons.} compared to the equivalent experiment having two predictors.

Additionally, we consider a technique often used by library authors to explicitly signal a break in ABI- a change in the file name reflecting the corresponding change in the library version number as defined in the libtools guidelines \cite{libtool}. As such, we further decompose our analysis into the set of libraries that changed their file name (after complete symlink resolution) between OS versions and those that did not. We find between $6$ and $20\%$ of binaries have a different file name across OS versions in the two-predictor case and between $2$ and $4\%$ in the three-predictor case. Finally, 72 linker scripts \cite{noauthor_undated-pj} mistaken for libraries were excluded.



\subsection{Predictor Agreement}

With the datasets in place, we turn our attention to the central question of agreement between predictors. For the cases with two predictors, at least one of them reports an ABI incompatibility in $63.3^{70.9}_{50.5}\%$ of cases when the file name changes and in $12.3^{16.0}_{9.0}\%$ of cases when it doesn't, and both predictors agree on the question of compatibility $58.9^{61.7}_{55.4}\%$ and $82.7^{86.2}_{78.5}\%$ of the time, respectively. For the three-predictor cases, the number of reported incompatibilities increases to $75.1^{92.3}_{60.8}\%$ for changed file names and $16.6^{14.0}_{19.5}\%$ for unchanged ones. Three-way agreement between the predictors is $56.1^{76.9}_{35.3}\%$ and $79.8^{83.4}_{76.5}\%$, respectively. To better understand these levels of disagreement, we dissect each case in detail.

In Table \ref{tab:twp_counts_changed}, we see $40.2\%$ (N=392) of libraries across all OS comparisons with changed file names are labelled as compatible by \textit{symbols} but as incompatible by Libabigail. The presence of such cases is expected as there are many other types of ABI incompatibilities beyond just changes to exported symbols. If we assume that a change in file name is used by the library's author to signal a break in ABI, then this behavior is substantiated by the tools in $86.1\%$ of cases. The remaining 135 cases where no incompatibility is detected represent either incompleteness in the tools or that a change in file name indicates something other than an ABI break (e.g., semantic versioning). When the file name does not change, Table \ref{tab:twp_counts_unchanged} shows that the fraction of incompatibilities detected only by Libabigail drops to $17.2\%$ but skyrockets to $83\%$ of all detected incompatibilities. Finally, we see $79.3\%$ (N=5274) of libraries have no detected incompatibilities which is consistent with our expectation given our assumption of the meaning of a change in file name.

\begin{table}
	\caption{\label{tab:twp_counts_changed}Prediction counts for the two-predictor case when file names changed.}
	\begin{tabular}{lcc}
    & \multicolumn{2}{c}{\textit{symbols}}\tabularnewline
        \cline{2-3} \cline{3-3} 
    Libabigail & compatible & incompatible\tabularnewline
    \multicolumn{1}{r}{compatible} & 135 & 0\tabularnewline
    \multicolumn{1}{r}{incompatible} & 392 & 447\tabularnewline
    \hline
  \end{tabular}
\end{table}

\begin{table}
	\caption{\label{tab:twp_counts_unchanged}Prediction counts for the two-predictor case when file names did not change.}
	\begin{tabular}{lcc}
    & \multicolumn{2}{c}{\textit{symbols}}\tabularnewline
        \cline{2-3} \cline{3-3} 
    Libabigail & compatible & incompatible\tabularnewline
    \multicolumn{1}{r}{compatible} & 5274 & 0\tabularnewline
    \multicolumn{1}{r}{incompatible} & 1143 & 233\tabularnewline
    \hline
  \end{tabular}
\end{table}

Looking now at the corresponding three-predictor cases, Table \ref{tab:thrp_counts_changed} shows that Libabigail and ABI Laboratory are in agreement $73.4\%$ of the time for libraries with changed file names have have only a single case where ABI Laboratory reports a detected incompatibility that Libabigail does not. Comparing against the \textit{symbols} predictor, the two are in agreement $74.3\%$ of the time. Quite surprisingly, \textit{symbols} reports 7 incompatibilities that ABI Laboratory does not. For the cases of libraries with no change in file name, shown in Table \ref{tab:thrp_counts_unchanged}, the agreement percentages increase to $95.6\%$ and $82.6\%$, respectively. This time, ABI Laboratory reports 63 incompatible libraries not labelled as such by Libabigail, and \textit{symbols} reports one such case.

As noted above, we encountered many difficulties while executing ABI Laboratory, and the technical issues continued as we attempted to retrieve its reports for the cases where it disagrees with the other two predictors. The tool merely reported that it was unable to find tens (sometimes hundreds) of unnamed symbols. We carefully examined the inputs and found no issues. In particular, the same libraries were readily processed by the other tools. At this time, we are unable to say anything further about the types of checks ABI Laboratory uses and how they differ from the other tools presented here.

\begin{table*}
	\caption{\label{tab:thrp_counts_changed}Prediction counts for the three-predictor case when file names changed.}
	\centering
	\begin{tabular}{lllll}
		\toprule
         & \multicolumn{2}{|c|}{\textit{symbols}} & \multicolumn{2}{|c|}{ABI Laboratory} \\ 
        \midrule
        \textbf{Libabigail} & compatible & not comp. & compatible & not comp. \\
        compatible & 5 & 0 & 4 & 1 \\
        not comp. & 41 & 63 & 28 & 76 \\
        \textbf{ABI Laboratory} & \multicolumn{4}{c}{} \\ 
        compatible & 25 & 7 & \multicolumn{2}{c}{}  \\
        not comp. & 21 & 56 & \multicolumn{2}{c}{} \\
        \bottomrule
	\end{tabular}
\end{table*}

\begin{table*}
	\caption{\label{tab:thrp_counts_unchanged}Prediction counts for the three-predictor case when file names did not change.}
	\centering
	\begin{tabular}{lllll}
		\toprule
         & \multicolumn{2}{|c|}{\textit{symbols}} & \multicolumn{2}{|c|}{ABI Laboratory} \\ 
        \midrule
        \textbf{Libabigail} & compatible & not comp. & compatible & not comp. \\
        compatible & 3075 & 0 & 3012 & 63 \\
        not comp. & 747 & 181 & 114 & 814 \\
        \textbf{ABI Laboratory} & \multicolumn{4}{c}{} \\ 
        compatible & 3125 & 1 & \multicolumn{2}{c}{}  \\
        not comp. & 697 & 180 & \multicolumn{2}{c}{} \\
        \bottomrule
    \end{tabular}
\end{table*}

\subsection{Filename and SONAME Changes}

Returning back to the two-predictor cases, there are clearly instances when the file name change (or lack thereof) does not reflect our assumed meaning. The libtools guidelines provide a second indicator for this purpose: the SONAME. This is a simple string containing the semantic version of the library based on a major, minor, and patch level (e.g., 3.2.1). For library authors that use libtools to build their projects, the SONAME is intended as an automatic means of changing the generated file name of the library. Of the tools we explore, only Libabigail uses the SONAME as a criterion for ABI compatibility- where a change in SONAME is considered an ABI break. We explore this for the two-predictor case in Table \ref{tab:twp_soname}. We see that a change in the file name and a change in SONAME are in agreement $57.5\%$ of the time. However, the SONAME is changed if and only if the file name is changed in just $27.4\%$ of cases. There are only a few libraries labelled as incompatible solely because of an SONAME change: 45 ($4.6\%$) when the file name changed and 48 ($0.7\%$) when it didn't.

\begin{table}
	\caption{\label{tab:twp_soname}Change in SONAME for libraries with two predictors.}
	\begin{tabular}{lcc}
    & \multicolumn{2}{c}{SONAME}\tabularnewline
        \cline{2-3} \cline{3-3}
    File name & changed & unchanged\tabularnewline
    \multicolumn{1}{r}{changed} & 230 & 609\tabularnewline
    \multicolumn{1}{r}{unchanged} & 52 & 1324\tabularnewline
    \hline
  \end{tabular}
\end{table}

We are able to correlate the many other ABI incompatibility checks Libabigail performs with the change in SONAME. Table \ref{tab:libabigail-detailed-results} lays out these results. We note that the percentages represent the fraction of reported breakages containing at least one of the features. Because a library may have many detected incompatibilities, these fractions do not add to $100\%$. For example, $52.6\%$ of all reported incompatibilities in libraries having both the file name and SONAME changed have \textit{at least} one function removed, but could also have functions with changed signatures. From the summary column, we see that, unsurprisingly, function signature changes dominate the space of reported ABI incompatibilities. Unexpectedly, modifications of enumerator values is the second-largest category followed by global variables and then virtual tables (vtable). The dearth of results for libraries without a change in file name but \textit{with} a change in SONAME is due to $92.3\%$ (N=48) of them being labelled as incompatible solely because of a change in SONAME.

\begin{table*}
    \begin{centering}
    \caption{\label{tab:libabigail-detailed-results}Frequency of ABI breakage types detected by Libabigail for two-predictor cases.}
    \begin{tabular}{c|l|r|r|r|r|r}
        \multicolumn{2}{c}{} & \multicolumn{4}{c}{} & \tabularnewline
        \cline{3-6} \cline{4-6} \cline{5-6} \cline{6-6} 
        \multicolumn{2}{c|}{} & \multicolumn{2}{c|}{Filename Change} & \multicolumn{2}{c|}{Filename No Change} & \tabularnewline
        \multicolumn{2}{c|}{} & \multicolumn{2}{c|}{} & \multicolumn{2}{c|}{} & \tabularnewline
        \cline{3-6} \cline{4-6} \cline{5-6} \cline{6-6} 
        \multicolumn{1}{c}{Feature} &  & \multicolumn{1}{r}{SONAME Change} & No Change & \multicolumn{1}{r}{SONAME Change} & No Change & Total\tabularnewline
        \hline 
        \multirow{4}{*}{functions} & removed & 52.6\% & 32.0\% & 7.7\% & 48.9\% & 43.7\%\tabularnewline
        \cline{2-7} \cline{3-7} \cline{4-7} \cline{5-7} \cline{6-7} \cline{7-7} 
         & changed & 48.7\% & 80.5\% & 0 & 28.0\% & 43.9\%\tabularnewline
        \cline{2-7} \cline{3-7} \cline{4-7} \cline{5-7} \cline{6-7} \cline{7-7} 
         & $\Delta$subtype & 39.6\% & 64.2\% & 0 & 6.6\% & 25.7\%\tabularnewline
        \cline{2-7} \cline{3-7} \cline{4-7} \cline{5-7} \cline{6-7} \cline{7-7} 
         & $\Delta$return type & 30.0\% & 51.9\% & 0 & 15.0\% & 26.3\%\tabularnewline
        \hline 
        \multirow{2}{*}{vtable} & added & 18.3\% & 25.0\% & 0 & 0.3\% & 8.9\%\tabularnewline
        \cline{2-7} \cline{3-7} \cline{4-7} \cline{5-7} \cline{6-7} \cline{7-7} 
         & removed & 16.5\% & 12.2\% & 0 & 0.3\% & 5.2\%\tabularnewline
        \hline 
        \multirow{3}{*}{enumerator} & added & 27.8\% & 7.9\% & 0 & 11.3\% & 11.8\%\tabularnewline
        \cline{2-7} \cline{3-7} \cline{4-7} \cline{5-7} \cline{6-7} \cline{7-7} 
         & removed & 25.7\% & 4.3\% & 0 & 11.0\% & 10.4\%\tabularnewline
        \cline{2-7} \cline{3-7} \cline{4-7} \cline{5-7} \cline{6-7} \cline{7-7} 
         & changed & 26.5\% & 8.0\% & 0 & 11.6\% & 11.9\%\tabularnewline
        \hline 
        \multirow{2}{*}{global variables} & removed & 19.6\% & 6.7\% & 7.7\% & 12.6\% & 11.6\%\tabularnewline
        \cline{2-7} \cline{3-7} \cline{4-7} \cline{5-7} \cline{6-7} \cline{7-7} 
         & changed & 7.8\% & 3.9\% & 0 & 5.4\% & 5.1\%\tabularnewline
        \hline 
    \end{tabular}
    \\ Note: We use the shorthand $\Delta x$ to represent ``a change in x'' to save space.
    \par\end{centering}
\end{table*}

\begin{table*}
    \centering
    \caption{\label{tab:comparison-with-abi-lab}Comparison with ABI laboratory}
    \begin{tabular}{|l|l|r|r|}
        \hline 
        \multicolumn{2}{|l|}{\textbf{Kind of ABI Change Detected}} & \textbf{ABI Lab} & \textbf{This Work}\tabularnewline
        \hline 
        \hline 
        \multicolumn{2}{|l|}{Removing functions from the library} & 55.67\% & 12.7\%\tabularnewline
        \hline 
        \multicolumn{2}{|l|}{Changing virtual table structure} & 1.38\% & 2.6\%\tabularnewline
        \hline 
        \multicolumn{2}{|l|}{Changing number or order of parameters} & 2.07\% & \multirow{3}{*}{12.8\%}\tabularnewline
        \cline{1-3} \cline{2-3} \cline{3-3} 
        \multirow{2}{*}{Changing type of parameter} & destructive & 9.55\% & \tabularnewline
        \cline{2-3} \cline{3-3} 
         & non-destructive & 15.89\% & \tabularnewline
        \hline 
        \multirow{2}{*}{Changing type of return value} & destructive & 0.12\% & \multirow{2}{*}{7.6\%}\tabularnewline
        \cline{2-3} \cline{3-3} 
         & non-destructive & 3.27\% & \tabularnewline
        \hline 
        \multicolumn{2}{|l|}{Changing parameter subtype} &  & 7.5\%\tabularnewline
        \hline 
        \multicolumn{2}{|l|}{Adding/removing ``static'' specifier} & 0.15\% & \tabularnewline
        \hline 
        \multicolumn{2}{|l|}{Changing values in enumeration types or macros} & 7.47\% & 3.6\%\tabularnewline
        \hline 
        \multirow{2}{*}{Global variables} & removed &  & 4.0\%\tabularnewline
        \cline{2-4} \cline{3-4} \cline{4-4} 
         & changed type &  & 1.5\%\tabularnewline
        \hline 
    \end{tabular}
\end{table*}

\subsection{Breakage Types}

While it was out of scope for our study to do a detailed categorization of types of breakages, we can classify them into the categories based on whether features (e.g., functions, vtable, enumerator, global variables) were removed, changed, or added \ref{tab:libabigail-detailed-results}. For each of these cases, we can show example cases for illustrative purposes.

\subsubsection{Function parameters}

An added or removed function speaks for itself as it means a symbol "mangled string" in C++ is either missing after a library change or newly present. As previously shown , a simple change
As shown previously \ref{example:function-parameter-add}, a function change that would lead to a breakage can be as simple as changing a parameter type. 

\paragraph{Structure layout change}

This kind of change might not be obvious on the level of the function, but could lead to a breakage. As an example, imagine we have a function that takes a structure S with members of an int and a double, and used by function foo:

\begin{lstlisting}
struct S { int x; double d; }
void foo(S s);
\end{lstlisting}

If we were to then change the structure S to instead have a char p (as shown below), although the signature of foo does not change, the structure itself would be a different size with an expectation of different parameter types. Passing the wrong structure type to either the new or old function would come down to passing a pointer when a double is expected, or vice versa. In both cases the application would not function as expected.

\begin{lstlisting}
struct S { int x; char *p; }
void foo(S s);
\end{lstlisting}

\paragraph{Subtype change}

This example is similar to the previous, however there is a layer of indirection through the inheritance of the Base class.

\begin{lstlisting}
struct Base { int x; };
struct Derived : Base {};
void foo(Derived *d) {}
\end{lstlisting}

In this example, if we were to change the Base class member from an int to a double, it would also not show up in the function signature for foo, nor would it show up in the Derived structure. This would lead to undefined behavior.

\begin{lstlisting}
struct Base { double x; };
struct Derived : Base {};
void foo(Derived *d) {}
\end{lstlisting}

\paragraph{Return type change}

We start with a structure S that is returned by a function foo.

\begin{lstlisting}
struct S { int x; double d; }
S foo();
\end{lstlisting}

We later change that return type to be an int. Since return types are not encoded in mangled strings, we would not see this change on the level of the symbol. It would also lead to undefined behavior.

\begin{lstlisting}
int foo();
\end{lstlisting}

\subsubsection{Virtual tables}

\paragraph{Adding a virtual function entry}

We start with a structure Base that has one entry in its virtual table, a function named bar.
Another function, foo takes a pointer to a Base as the only parameter.

\begin{lstlisting}
struct Base { virtual void bar(); };
void foo(Base *b) {}
\end{lstlisting}

Now let's imagine that base has another entry added to its virtual table -- a function named baz.
Notice that the function foo that uses the Base has superficially not changed.

\begin{lstlisting}
struct Base {
	virtual void bar();
	virtual void baz();
};
void foo(Base *b) {}
\end{lstlisting}

This is a breaking change because virtual tables are part of the layout of the structure Base, and the function is now expecting a different layout that might not be obvious. The order in which virtual functions are added to the table is not specified, meaning the compiler can choose to do what it wants. This means that the original entry for function bar might result in a different location.

\subsubsection{Enumerator}

\paragraph{Adding an enumerator}

We start with an enum, Foo, that has two enumerators, and that is used by a function foo as the only parameter.

\begin{lstlisting}
enum class Foo {ONE,TWO};
void foo(Foo f);
\end{lstlisting}

If we then add a third enumerator to Foo, the function signature does not change, however the underlying data that is expected has. This would lead to undefined behavior. This is another kind of layout change, however distinct because enumerations do not have sizes. An enumerator is a label that contains a value, which is an implementation detail.

\begin{lstlisting}
enum class Foo {ONE,TWO,THREE};
void foo(Foo f);
\end{lstlisting}

This means that we can tell the enumerations to take on certain values, as we do below.

\begin{lstlisting}
enum class Foo {ONE=17,TWO,THREE};
void foo(Foo f);
\end{lstlisting}

While this might not be an ABI error in the traditional sense, this would be a behavioral change, akin to an integer parameter taking on different values. It would be ultimately important to signal these kinds of behavioral changes, regardless of how they are classified.

\subsubsection{Global Variables}

\paragraph{Changing a global variable}

While the classification of global for a variable indicates visibility, the linkage of the variable means that it is part of the ABI surface. In layman's terms, a library might expect to use a global variable 
provided by another one. In the example below, both x and y have global visibility. However, in adding static to y, this means it has internal linkage and can no longer be seen by the ABI surface. 

\begin{lstlisting}
int x=1;
static int y=2;
\end{lstlisting}

Both of these would be called global variables, however the second would not be seen in the context of an ABI. In that int x is seen, a change to its type would be breaking. The kinds of changes previously discussed in the previous sections could also apply here.

\section{Discussion}
\label{sec:discussion}

We conducted an experiment to compare tool predictions across several versions of the Fedora operating system, focusing on comparing the strengths and weaknesses of the tools. Although we've done a quantitative analysis, we are notably still doing a qualitative comparison of these tools.
Our results are limited based on the dearth of tools in this space, and as a result, we recognize that much of our qualitative analysis ultimately focused on Libabigail, which is arguably the most mature tool in the space.

Our decision to focus on the two-predictor case is not only driven by having 49.5\% more results, but also a desire to focus on the tool that is most widely used in developer communities (e.g., Google's Android team and Red Hat Linux). In stark contrast, at the time of this writing, the ABI Laboratory abi-compliance-checker \cite{noauthor_undated-jk} has not been updated on the GitHub repository for over a year. We believe there is valuable knowledge embedded in the tool, and hope that either the maintainer can continue working on it, another set of maintainer(s) can contribute, or the knowledge can be extracted and moved elsewhere. It is not clear the extent to which the community or original developers are interested in such an effort.
 
We found that, on average, about 12$\%$ of libraries have a different file name between OS versions and account for $37.9\%$ of all libraries with a least one ABI break reported. Interestingly, there is nearly uniform N-way agreement between the predictors when looking across the subsets of changed and unchanged ($\sim 75\%$) file names. Recall that the \textit{symbols} predictor only analyzes the sets of exported symbols (i.e., those in .symtab and .dynsym ELF sections). This means that the cases where the file name does not change, the simple symbol name check discovers $\sim 75\%$ of libraries that would be labelled as incompatible by the more in-depth analyses performed by Libabigal and ABI Laboratory. Additionally, we found the \textit{symbols} predictor to be about a thousand times faster than the other two with its execution on the order of a few milliseconds compared to the other two on the order of a few seconds. This suggests that a simple symbols comparison could both be an effective (but not complete) and fast means of examining a very large body of libraries like a complete Linux distribution in near real-time.

Given an approximately 80\% of agreement of the symbols predictor with Libabigail when the file name changes, our results suggest that the majority of cases are related to symbol names. Our results suggest that SONAME and filename changes are not reliable to indicate change, as it can go either way to have a library with a changed name that has not actually changed, or a library with an unchanged name that has. We suspect that the underlying SONAME is perhaps overlooked by developers, the actual practice is being deprecated in favor of using a file name, or that developers are treating the SONAME differently than as defined in the libtool guidelines. Given this result, it appears that software today does not reliably use the SONAME to indicate an ABI breaking-change, and thus we suggest that the SONAME is not a good indicator of such a break. Between the two, SONAME is slightly less reliable than filename. A better alternative, given no tool to do an assessment, might be to look at symbols. We suggest to the interested developer that needs to assess ABI breaks across a large set of pairs (where computational complexity and running time is an issue) that a simple comparison of symbol sets can get them most of the way there. Given the large speed difference (three orders of magnitude faster) between the simple symbols predictor and the others, the developer can detect a large number of problems substantially faster.

In comparing our results to previous work (see Table \ref{tab:comparison-with-abi-lab}) this previous work \cite{5952657} found that approximately 55\% of breakages were due to removing functions in a library, whereas we found approximately 12\%. We believe this discrepancy could result from a difference in the size of datasets used for the analysis. The previous work used a dataset of approximately 250, and our dataset size is a factor of 27 times that. We are interested in comparing the result that inspects destructive vs. non-destructive types of changes, however while our primary tool Libabigail has that information, it is not presented in a machine-readable way. A substantially useful addition to the Libabigail software would be to better expose results in a machine-readable form. Both our analyses and previous work detect enumerator breakages, however our work detects approximately half the quantity, percentage wise. The also could be attributed to the size differences of the datasets. As both results suggest this is the second most abundant type of change, our findings suggest that enumerators are an important place where ABI checks should be performed.

The third most common type of breakage, as reported by Libabigail, are from global variables (approximately 5.5\%). We find it interesting that this breakage is not detected by ABI Laboratory. This case represents an example of how tools choose to create an internal model of ABI that is likely to differ with other tools. Work to better standardize the underlying model that tool implementations use would be a suitable contribution to this space. Finally, the fourth most common breakage for both analyses is with respect to virtual tables. We find a higher fraction of libraries that exhibit changes in the structure of these tables (~7.5\% of all libraries) than the previous analysis, and we are also able to detect changes in derived types in a type hierarchy (e.g., changing parameter subtype), which was not stated in the previous analysis. Because the previous work was based on source-level analysis, they were able to directly detect the addition or removal of the \textit{static} specifier functions. This type of change would manifest in a binary analysis either as a added/removed function symbol or a add/removed function parameter (via the hidden \textit{this} parameter in C++).

\subsection{Limitations}

Although we have control in the compiler and general environment when using a container for our experiments, we don't have control over the microarchitecture of the public GitHub runners. In general, although there are known ABI issues when changing underlying microarchitecture (Section 3.2.3 of \cite{noauthor_undated-ni}), to our knowledge, the current tooling available is not able to check at this level of detail, so it would not impact our superficial analysis. For provenance, the runners that we encountered included broadwell, cascade lake, haswell, and generic x86\_64\_v4 (unspecified). 

We recognize that our results might be biased based on using a production-ready set of binaries provided by a single package manager for the Fedora operating system. We suspect that looking at binaries from a different package manager or Linux distribution could result in different results. Finally, our symbols analyses assume binaries compiled on a Unix type system. The tools that we chose are also stated to be primarily intended for ELF binaries. Extending this work to other types of binaries or a Windows system would require implementing this process for a different binary format or algorithm, in the case of Windows.

\section{Conclusion}
\label{sec:conclusion}

In this paper we have reviewed well-known tools for checking ABI compatibility of software. As the number of architectures, compilers, and languages increases, having an ability to model ABI compatibility is becoming increasingly important, along with having automated means to assess it. We are pleased to share this early work that compares agreement between ABI prediction tools, and hope that it inspires other developers to contribute to work in space.

Although ABI is a niche topic of interest in the developer space, we strongly recommend that developers working on C/C++ packages add basic ABI checks to their continuous integration as a good practice. To support this, we have created a GitHub action \cite{libabigail-action} that easily runs Libabigail, and allows the developer to check some number of libraries in development (in a pull request) against a previous latest release, and the current main branch. While there aren't many tools in the space, we champion using a tool like Libabigail, and specifically for this tool, hope to see continued development toward having machine-readable output to allow for easier interpretation of the results. While ABI breaks are not avoidable, we believe these checks to be important so developers have awareness about updates to their software that might break ABI.

We are concerned about the incredibly small number of projects working in this space given how fundamental this problem is in modern software. Application binary interfaces changing and resulting in an error that can hinder usage, performance, or portability is an inescapable problem. Having these kinds of tools embedded in build tools would be incredibly beneficial to the process. If this kind of work is valuable or essential for development (and we suspect it is the case that there are far more wanting to use this kind of tooling than develop it) there is a huge opportunity to better form a community focused around shared goals for development. Based on our observation of likely different underlying models between the tools that we tested, we suggest moving forward that work is needed in this space to better categorize and communicate more complex ABI breakages, both for understanding and mapping of the space. 

\subsubsection*{Contributions}

Author VS designed the original experiment, implemented the spliced software framework and automation (containers and workflows along with software) to interact with all predictors, ran the experiments, and wrote the first and second drafts of the manuscript.
Author TH contributed significantly to the experiment design, outline for the paper, final draft, and analysis and interpretation of results.

\subsubsection*{Acknowledgements}

We are thankful to Ben Woodard, Jim Kupsch, and Bart Miller for feedback on an initial draft, and specifically to Ben Woodard and Dodji Seketeli for their work on Libabigail. Thank you also to Matt Legendre and Todd Gamblin for providing a funding structure to support this work via the BUILD SI project.
Finally, thank you to "Old Man Cat," "Tiny Sir", "Enzo" who was consistently present for pair programming and work sessions, and also was consistently no use at all. 

This work was performed under the auspices of the U.S. Department
of Energy by Lawrence Livermore National Laboratory, LLNL-JRNL-842973-DRAFT.

\bibliographystyle{unsrtnat}

\bibliography{references}  






\end{document}